%
\input harvmac.tex
\hfuzz 15pt
\input amssym.def
\input amssym.tex
\input epsf\def\tfig#1{{
\xdef#1{Fig.\thinspace\the\figno}}Fig.\thinspace\the\figno
\global\advance\figno by1}


\input epsf

%


\def\({\big(}
\def\){\big)}

\def\pd{\partial}

\def\inv{^{-1}}

 \def\frac#1#2{ {{\textstyle{#1\over#2}}}}
\def\inv{^{\raise.15ex\hbox{${\scriptscriptstyle -}$}\kern-.05em 1}}

\def\a{\alpha}

\def\g{\gamma}
\def\d{\delta}

\def\l{\lambda}

\def\et{\eta}
\def\vr{\varrho}
\def\s{\sigma}

\def\G{\Gamma}
\def\D{\Delta}

\def\P{\Pi}

\def\O{\Omega}

\def\no{\noindent}

\def\tri{\triangle}

\def\rb{ \noindent $\bullet$\ \ }

\def\zu{\Upsilon_b}

\def\IC{{ \Bbb C} }

\def\dC{C\kern-6.5pt I}

       \def\CN{{\cal N}}

\def\bz{\bar z}
\def\bx{\bar x}

\def\bg{\bar \g}
\def\bu{\bar u}

\input amssym.def
\input amssym.tex
\def\IC{\Bbb C}


  \def\zT{\Theta} 
 \def\H_{H_{1,2}}  \def\zT_{\Theta_{1,2}}
 \def\O_{O_{1,2}} \def\bH_{{\bar H}_{1,2}} 
  
 \def\V_{V_{1,2}} 
 \def\D_{D_{1,2}} \def\bD_{{\bar D}_{1,2}}


       \def\cD{{\cal D}}

     \def\cD_{{\cal D}_{1,2}} 
 \def\bcD_{{\bar {\cal D}}_{1,2}}


   \def\dC{I\!\!\!\!C}




\def\ho{{1\over 2}}

\def\dal{
\vbox{
\halign to5pt{\strut##&
\hfil ## \hfil \cr
&$\kern -0.5pt
\sqcap$ \cr
\noalign{\kern -5pt
\hrule}
}}\ }



\def\un{{\bf 1}}



\def\adtmp#1#2#3{{\it Adv. Theor. Math. Phys. } {\bf  #1} (#2) #3}

\def\lmp#1#2#3{{\it Lett. Math. Phys.} {\bf  #1} (#2) #3}

\def\npb#1#2#3{{\it Nucl. Phys.} {\bf B #1} (#2) #3}

\def\plb#1#2#3{{\it Phys. Lett.} {\bf B #1} (#2) #3}

\def\rb{$\bullet\ \ $}

\lref\LMRS{ S. Lee, S. Minwalla, M. Rangamani and N. Seiberg, Three point functions of chiral operators in $D=4, \CN=4$ SYM at large $N$, \adtmp{2}{1998}{697}
hep-th/9806074.}
\lref\FMMR{D. Freedman, S. Mathur, A. Matusis and  L. Rastelli, 
Correlation functions in the CFT$_d$/AdS$_{d+1}$ correspondence, \npb{546}{1999}{96}, 
hep-th/9804058.}
\lref\Th{ C. Thorn, Liouville perturbation theory, {\it Phys. Rev.} {\bf D 66}, 027702 (2002), hep-th/0204142.}
\lref\D{V.K. Dobrev, Intertwining operator realization of the AdS/CFT correspondence, Nucl. Phys. {\bf B 553} (1999) 559, 
hep-th/9812194.}
\lref\ZZ{A. Zamolodchikov and Al. Zamolodchikov, Structure constants and conformal bootstrap in Liouville field theory, 
\npb{477}{1996}{577}, hep-th/9506136.}
\lref\DO{ H. Dorn, H.J. Otto: Two and three point functions in
Liouville theory,  \npb{429}{1994}{375}, hep-th/9403141.  }
\lref\MT{ P. Menotti and E. Tonni, hep-th/0508240.}
\lref\P{A.M. Polyakov, Gauge Fields and Space-Time, 
{\it Int. J. Mod. Phys.}  {\bf A17} S1 (2002) 119, hep-th/0110196.}
\lref\AT{A. A. Tseytlin,  On semiclassical approximation
and spinning string vertex operators in $AdS_5\times S^5$, \npb{664}{2003}{247},  hep-th/0304139.}
\lref\Gaw{K. Gaw\c{e}dzki, Non-compact WZW conformal field theories,  in:  Proceedings  of  NATO ASI  Cargese 1991, "New symmetry  principles in quantum field theory",  eds. , J. Froehlich, G. {}'T Hooft, A. Jaffe, G. Mack, P.K.Mitter, R. Stora, Plenum Press (1992) p. 247, hep-th/9110076.}
\lref\BORT{ J. de Boer, H. Ooguri, H.  Robins, J. Tannenhauser, String Theory on AdS$_3$, 
 {\it JHEP}  {\bf 9812} (1998) 026, arXiv:hep-th/9812046. }
 \lref\KL{ T. Klose and T. McLoughlin, A light-cone approach to three-point  functions in AdS$_5 \times S_5$,
arXiv:1106.0495.}
 \lref\JW{R.  Janik and  A. Wereszczynski, Correlation functions of three heavy operators - the AdS$_5$ contribution, 
 {\it JHEP}   {\bf 1112}  (2011) 095, arXiv:1109.6262.}
 %
 %
%
%
\lref\GabKir{M.R. Gaberdiel and  I. Kirsch, Worldsheet correlators in AdS(3)/CFT(2), 
{\it JHEP}  {\bf 0704} (2007), 050,  hep-th/0703001.} 
\lref\DalPak{ A. Dabholkar and A.  Pakman, Exact chiral ring of AdS$_3$/CFT$_2$,  {\it Adv. Theor. Math. Phys.} {\bf 13} (2009) 409, 
arXiv:hep-th/0703022. }
\lref\PakS{A.  Pakman and A.  Sever, Exact N=4 correlators of AdS$_3$/CFT$_2$
 Phys.Lett. {\bf B652} (2007) 60, arXiv:0704.3040.}
 \lref\GM{P. Ginsparg and  G. Moore, Lectures on 2D gravity and 2D string theory, hep-th/9304011.}
 \lref{\Tea} { J. Teschner,  Operator product expansion and factorization in the $H_3^+$-WZNW model,
 \npb{546}{1999}{390},  hep-th/9906215.}
 \lref\Te{ J. Teschner, On structure constants and fusion rules in the $SL(2,\IC)/SU(2)$ WZNW model
\npb{546}{1999}{390}, hep-th/9712256.}

\lref\FGPP{ FGPP }
\lref\ZF{V. A. Fateev and A. B. Zamolodchikov,  Operator algebra and correlation functions of the two-dimensional Wess-Zumino $SU(2)\times SU(2)$ chiral model,  {\it Sov. J. Nucl. Phys.}  {\bf 43} (1986) 657.}
\lref\DF{ Vl. S. Dotsenko and V.A. Fateev, 
Conformal algebra and multipoint correlation functions in 2D statistical models,
\npb{240}{1984} {312}.}
\lref\KZ{V.G. Knizhnik and A.B. Zamolodchikov, 
 Current algebra Wess-Zumino model in two dimensions, \npb{247}{1984}{83}.}
\lref\FGP{
 J.L. Petersen, J. Rasmussen and M. Yu,
   Fusion, crossing and monodromy in conformal field theory
   based on $sl(2)$ current algebra with fractional level, \npb{481}{1996}{577}, 
   hep-th/9607129; 
   P. Furlan, A.Ch. Ganchev and V.B. Petkova,
 $A_1^{(1)}$ admissible representations - fusion transformations
and local correlators,  \npb{491}{1997}{635}, hep-th/9608018.}
\lref\MalO{J. Maldacena and  H. Ooguri,  Strings in AdS$_3$  and the SL(2,R)  WZW Model. Part 1: The spectrum, 
{\it J. Math. Phys.} {\bf 42} (2001) 2929, hep-th/0001053.}
\lref\MalOb{J. Maldacena, H. Ooguri and J Son, 
Strings in AdS$_3$ and the SL(2,R) WZW Model. Part 2: Euclidean Black Hole, hep-th/0005183.}
\lref\HMW{
D. Harlow, J. Maltz and  E. Witten, Analytic Continuation of Liouville Theory, arXiv:1108.4417.}
\lref\AdSCFT{ J. M. Maldacena, The large N limit of superconformal
field theories and supergravity'', \adtmp{2}{1998}{231}, hep-th/9711200.}
\lref\GKP{
S. S. Gubser, I. R. Klebanov
and A. M. Polyakov, Gauge theory correlators from non-critical
string theory, \plb{428}{1998}{105}, hep-th/9802109.}
\lref\EW{
E. Witten, Anti-de Sitter space and holography, 
\adtmp{2}{1998}{253}, arXiv:hep-th/9802150.}
\lref\Rev{ N. Beisert, Ch. Ahn, L.F. Alday, Z. Bajnok, J.M. Drummond,
L. Freyhult, N. Gromov, R.A. Janik, V. Kazakov,
T. Klose, G.P. Korchemsky, Ch.  Kristjansen, M.
Magro, T.  McLoughlin, J.A. Minahan, R.I. Nepomechie,
A. Rej, R. Roiban, Sakura Schafer-Nameki, Christoph Sieg,
Matthias Staudacher, A. Torrielli, A.A. Tseytlin, P.
Vieira, D. Volin, K. Zoubos, ``Review of AdS/CFT
Integrability: An Overview'', \lmp{99}{2012}{3}, 
arXiv:hep-th/1012.3982. }
\lref\BuchT{ E. I. Buchbinder and A. A. Tseytlin, On semiclassical approximation for correlators of closed string
vertex operators in AdS/CFT, {\it JHEP}  {\bf 1008}  (2010) 057,
arXiv:hep-th/1005.4516.}
\lref\Janik{ R. A. Janik, P. Surowka and A. Wereszczynski,
On correlation functions of operators dual to classical spinning
string states, {\it JHEP}  {\bf 1005}  (2010), 030, 
arXiv:hep-th/1002.4613.}
\lref\RT{ R.~Roiban and A.~A.~Tseytlin,
On semiclassical computation  of 3-point functions of closed
string vertex operators in $AdS_5\times S^5$,  {\it Phys. Rev.}  {\bf D82}
106011 (2010) arXiv:hep-th/1008.4921.}
\lref\KK{Y. Kazama, S. Komatsu, On holographic three point functions
for GKP strings from integrability, {\it JHEP}  {\bf  1201} (2012), 110, 
arXiv:hep-th/1110.3949.}
\lref\BT{ E. I. Buchbinder and  A. A. Tseytlin, Semiclassical correlators of three states
with large $S^5$ charges in string theory in $AdS_5 \times S^5$,
{\it Phys. Rev.} {\bf D85} 026001 (2012) arXiv:hep-th/1110.5621.}

\lref \Wiegmann{P. Wiegmann, Extrinsic geometry of superstrings, \npb{323}{1982}{330}.}

\lref\MTs{R.R. Metsaev and  A.A. Tseytlin, Type IIB superstring action in $AdS_5 \times  S^5$   background, 
          \npb{533}{1998}{109},  hep-th/9805028. }

\lref\KalTs{R. Kallosh and A.A. Tseytlin,  Simplifying superstring action on $AdS_5 \times S^5$, JHEP {\bf 10} (1998) 016, hep-th/9808088.}


\overfullrule=0pt

\Title{\vbox{\baselineskip12pt\hbox
{}\hbox{}}}
{\vbox{\centerline
 {On the semiclassical }
 \medskip
 \centerline{ 
3-point function  in  
 AdS$_3$}
 \vskip 0.8cm
 \bigskip
  \vskip1.5pt
}}

 \centerline{ P. Bozhilov$^{a}$, P. Furlan$^{b,c}$, V.B. Petkova$^{a}$ and  M. Stanishkov$^{a}$ }
\vskip 5pt
\medskip
\bigskip

   \centerline{ \vbox{\baselineskip12pt\hbox
{\it $^{a)}$  
Institute for Nuclear Research and Nuclear Energy, }
}}
 \centerline{ \vbox{\baselineskip12pt\hbox
 {\it Bulgarian Academy of Sciences, Sofia, Bulgaria}
 }}
 \vskip 5pt
\medskip
 \centerline{ \vbox{\baselineskip12pt\hbox
{\it $^{b)}$Dipartimento  di Fisica 
 dell'Universit\`{a} di Trieste, Italy,}
 }}
\centerline{ \vbox{\baselineskip12pt\hbox
{\it $^{c)}$Istituto Nazionale di Fisica Nucleare (INFN),
Sezione di Trieste, Italy,}
}}

\vskip 1.8cm
 

\noindent

\no 
We reconsider the problem of determining the semiclassical 3-point  function in the
Euclidean Ad$S_3$ model. Exploiting the  affine symmetry of the model we 
use  solutions of  the classical Knizhnik-Zamolodchikov (KZ) equation 
to compute the saddle point of the action in the presence of three vertex operators.
This alternative derivation     reproduces the "heavy charge" classical limit
of the quantum 3-point correlator. It is  different from the recently proposed 
expression obtained by generalised Pohlmeyer reduction in AdS$_2$. 


\bigskip
\bigskip\bigskip
\bigskip

\no
-------------------------------------------------------------
\medskip
\no
plbozhilov@inrne.bas.bg, furlan@ts.infn.it, petkova@inrne.bas.bg, marian@inrne.bas.bg
\Date{}


\newsec{Introduction}

The  AdS/CFT conjecture \AdSCFT,\GKP,\EW\  
implies   that the  correlation functions  in the dual   (boundary)   quantum field theory  can be computed alternatively  in   string theory,  i.e., essentially by the methods of  a  two - dimensional   theory.  
The first computations  however  were  mostly   performed  in 
the  supergravity approximation, representing the correlators in terms of 
 integrals over the target (bulk) coordinates  \FMMR, \LMRS. 

On the string side one may start with a  semiclassical
approach, when the string path integral for the correlation
functions   is evaluated in the saddle-point approximation
with  large  {}'t Hooft coupling  
 ${\lambda}>\!>1$. In this calculation one has to identify 
the correct vertex operators \P, \AT\ and to find the corresponding
classical solutions, which provide the appropriate saddle-point
approximation. 
Some preliminary results for the three point function of three heavy operators are already available
\JW, see also  \KL, \KK,  \BT.  
 Here
we propose to use our knowledge 
of the  quantum Liouville and
WZW theories 
to such  semiclassical computations.

Semiclassical considerations of the (euclidean) Ad$S_3$  string theory have been initiated e.g., in \BORT, where  classical solutions of the equations  of motion  in the absence of sources, or with one vertex insertion have  been constructed.
On the other hand one can compute  straightforwardly the classical limit of the known quantum 3-point OPE coefficients, i.e., for $b^2 \sim {1\over \sqrt{\l}}  \to 0$, and heavy charges $\tri_i=-2j_i= {2\eta_i\over b^2}$, s.t.,  $\eta_i$ are  finite; here $\tri$ stands for  the scaling dimension  in  the dual CFT.   
Such configurations dominate the  saddle point of the  action in the presence of  sources, i.e., classical vertex operators. 
The resulting expression in this limit  is similar to the semiclassical Liouville result  \ZZ, due to the fact, that the quantum 
(euclidean) AdS$_3$ 3-point function is expressed by a formula \Te, \Tea\ closely related to the Liouville one. 

This semiclassical 
limit of the quantum $AdS_3$ 3-point function  yields an expression which differs 
  from the recently proposed one \JW. The latter is   interpreted as the  Ad$S_2$  part of the correlator of three heavy strings propagating  in  AdS$_2\times S^5$ model,   and  is assumed 
to be universal for heavy (scalar)  $AdS_{d+1}$  
operators. 
This 
motivated us to reconsider the problem and compute directly the semiclassical constant, generalising the method
in \ZZ,    which was proposed originally as a semiclassical check of the 
quantum Liouville 
3-point constant.  
  While the quantum AdS$_3$ theory  and its applications to  the  
 superstring on $AdS_3 \times S^3\times M^4$  
  in the NS-NS background  (\GabKir, \DalPak\  and references therein)  is well studied, and, thus,  
this  is no more than a toy model 
 in the  semiclassical context under consideration,  the  elaboration 
 of  2d CFT techniques  is important for the analogous unsolved problems in  more realistic and less known cases. 

Our derivation is based on the affine algebra symmetry of the model generated by a current and (in the euclidean version) 
its complex conjugate. This leads to  a 
 chiral 
equation,  
a  classical version of the KZ equation \KZ.
 The equation determines  the classical fundamental vertex $V$ of isospin  $ j= 1/2$  as a function of the coordinates of the three vertex sources. The solution  is then used,  analogously to  \ZZ,   to evaluate  the contribution of the sources to the saddle point of the action and thus  
 to compute the semiclassical 3-point function, confirming the direct  classical limit of the quantum correlator.

\newsec{Summary of  the AdS$_3$ data}

In this mostly preliminary section we summarise some  basic data  on the non-compact $\hat{sl} (2)$ WZW model
 \Gaw,  \Te.

\no \rb 
 The Euclidean AdS$_3$  is  the  coset $\simeq SL(2,\IC)/SU(2)$ 
\eqn\hth{-X_{-1}^2+\sum_{i=1}^3X_i^2=-1 
}
parametrised  in $SL(2,\IC)$ as 
\eqn\param{
g(X) = X_{-1} \un_2- X_i \s_i
= \pmatrix{e^{-\phi} +|\g|^2e^{\phi}& e^{\phi}\g \cr e^{\phi}\bar{\g}& e^{\phi}}=
\pmatrix{1&\g\cr 0&1} \pmatrix{e^{-\phi}&0\cr 0&e^{\phi}} \pmatrix{1&0\cr \bar \g&1} \,,
}
$$
ds^2= {1\over 2} {\rm tr}(g^{-1}d g )^2= d\phi^2 +d\g\,d\bar{\g} e^{2\phi} \,.
$$

\medskip
\no
In the WZW classical action the coordinates $\phi(z,\bz)\,, \g(z,\bz)$ are  2d fields 
\eqn\clasact{
S_{AdS}
= {k\over \pi} \int d^2 z (\pd_z \phi \,\pd_{\bz}\phi +\pd_z \bar{\g}\, \pd_{\bz} \g e^{2\phi})\,.
}

\no
The classical equations  of motion 
\eqn\eqm{\eqalign{
&\pd_{\bz}\pd_z \phi  = e^{2 \phi} \pd_z \bar{\g}\, \pd_{\bz} \g\,, 
\cr
&\pd_z\( (\pd_{\bz} \g) e^{2 \phi}\)=0\,, \ \  
\pd_{\bz}\( (\pd_{z} \bg) e^{2 \phi}\)=0 
}}
imply that the currents 
\eqn\chcur{ 
J(z)= k\, \pd_z g g^{-1}=k\, \pmatrix{-\pd_z\phi +\g \pd_z\bg\,  e^{2\phi} &
 2 \g \pd_z\phi-\g^2 \pd_z\bg e^{2\phi} +\pd_z \g \cr
  \pd_z\bg e^{2\phi}  & \pd_z\phi -\g \pd_z\bg e^{2\phi} } =J^a t^a 
}
and $\bar{J}(\bz)=k\, g^{-1}{\bar \pd}_z  g$ are conserved (chiral), $ {\pd}_{\bz} J(z)=0\,,\  \pd_z \bar{J}(\bz)=0$
 and vice versa.
 
 \no
 In other words $g(z,\bz)$ satisfies two   chiral first order equations
\eqn\chire{k\,  \pd_z g= J(z) g=J(z)^a t^a g \,, \ \  k \,   \pd_{\bz}  g= g \bar{J}^a(\bz) t^a \,.
}

The translations in the diagonal action of  $SL(2,\IC)$ on the coset 
shift   $\g\to \g-x$. On the projected group element
\eqn\vert{
V=V_{j={1\over 2}} (z,\bz;x, \bar x)=e^{-\phi} +|\g-x|^2e^{\phi}=  (1, -x) g(z,\bz) \pmatrix{1\cr  -\bar x}
}
the generators $t^a$ of $sl(2)$  
are  realised by standard differential operators with respect to the isopin variable $x$. 
They are  determined from
$$
k\, \pd_z V=  (1, -x) t^{a} g(z) \pmatrix{1\cr  -\bar x}\,  J^{a}(z)= :J^{a}(z) D^{a}(x) V \,, \ 
$$
  and analogously $\bar{D}^{a}(\bar x)$ are determined from the right action of $t^a$.
General vertex operators   are given by
 \eqn\qvert{
V_j(z,\bz,x,\bar x) = 
(e^{-\phi(z,\bz)}+|\g(z,\bz)-x|^2e^{\phi(z,\bz)})^{2j}\,.
}

\no \rb 
In the quantum theory  $k \to k-2=1/b^2$ and furthermore a curvature term is added. 
In the AdS$_3/CFT_2$ correspondence \qvert\  is the kernel of the integral boundary-bulk operator
with boundary conformal dimension $\tri=-2j$. Its  "world sheet" (Sugawara) scaling  dimension  is
\eqn\sudim{
\delta^{Su}(j)= -b^2\, j (j+1)=\a(b-\a)=\delta^L(\a)-\a/b\,.
}
We have used the notation   $\a_i=-j_i b$  for the vertex charges  to compare with the 
Virasoro  theory of central charge 
$c>25$  (Liouville theory).  The two Virasoro theories with generic $c=13-6(b^2+{1\over b^2})  <1$ and $c=13+6(b^2+{1\over b^2}) >25$ 
can be realised via 
quantum Hamiltonian reduction of the $\hat{sl}(2)$ (respectively,  compact and non-compact)  WZW models.  Accordingly 
the 3-point WZW OPE constants 
 are closely related to the  Virasoro ones. 
 In particular,  in the  non-compact 
 case  the 3-point constant is given  \Te\ by  the DOZZ  Liouville expression  up to a simple 
 $\g(x)=\Gamma(x)/\G(1-x)$  factor
\eqn\adst{\eqalign{
C(j_1,j_2,j_3) & =
(\nu(b))^{1+j _{123}}\,
{\zu(b)\, 
\over
\zu(\a _{123}-b)} 
\prod_{i=1}^3 {\zu(2\a_i)\over \zu(\a_{123}-2\a_i)}\cr
&\sim 
\g((\a_{123}-Q){1\over b}) 
C_L(\a_1,\a_2,\a_3)
}}
where $\zu(x)=\zu(Q-x)$ 
 is expressed by  Barnes double Gamma functions,  $\nu(b)$ is an arbitrary constant and $\a_{123}=\a_1\!+\!\a_2\!+\!\a_3$.\foot{We restrict here to three spectrally unflowed representations, cf.  \MalO, \MalOb\  for the full spectrum of the model.} 
\medskip 

\no
\rb  
With  $k$   replaced by the shifted $k-2=1/b^2$ in the action \clasact\   the semiclassical limit corresponds to $b^2 \to 0$ and "heavy"  charges $j$
 \eqn\heav{
 j=-{\a \over b}= -{\eta\over b^2}\,, \et - {\rm finite}\,, \ \ 
b^2 \delta^{Su}(j) \to   -\eta^2 \,. 
}
In this limit, described for  the Liouville theory in \ZZ,  the 
 function $\zu$ goes to
\eqn\Fint{\eqalign{
\log \zu({\eta\over b})  \to 	{1\over b^2}F(\eta):&= {1\over b^2}\int_{1/2}^{\eta} dx \log \g(x)\cr
}}
so that  for the 3-point constant   \adst\  one obtains 
\eqn\quasit{\eqalign{
-b^2 \log C(-{\eta_1\over b^2},-{\eta_2\over b^2},-{\eta_3\over b^2})  \to &-b^2\log  C^{(cl)}(\eta_1,\eta_2,\eta_3)\cr
=(\eta_{123}) \log \nu(b) & +F(\eta_{123})-F(0) +\sum_i (F(\eta_{123}-2\eta_i)- F(2\eta_i))\,.
}}
The heavy charge classical limit describes the semiclassical 3-point function of  vertex operators    which
 is dominated by the saddle point of the classical action, i.e., on some solution of the classical equations of motion \eqm\ 
which depend on the charges and coordinates  of  the sources - the   vertex operators. 
Following and extending the  approach in  \ZZ\ in the Liouville theory we shall reproduce  in the next section
 formula \quasit\  by first  describing explicitly   these classical solutions and  then 
 directly  computing the semiclassical 3-point function. 
\medskip

\no
\rb Just for comparison recall the "light charge" classical limit of \adst: for $b\to 0$ consider    $2\a_i=\tri_i b$  with fixed  $\tri_i=-2j_i$. 
 In this limit \Th\
\eqn\light{
{\zu(b \, \sigma)\over \zu(b)} \to {b^{1-\sigma}\over \Gamma(\sigma )} 
}
and thus from  \adst\   one   reproduces, up to trivial field renormalisation,    the expression for the $AdS_3$ 3-point constant computed in the supergravity
approximation \FMMR.\foot{The formulae in  the semiclassical considerations in \KL,\BT\  correspond to  "light charge" classical limit, in which  all $\tri_i$ are  furthermore taken big,  exploiting the  Stirling formula for the asymptotics of the Gamma functions. 
In this  limit  of  the supergravity AdS$_{2d+1}$  constants,  as well as of  their $S^{2d+1}$ analogs,  the dependence 
on $d$ is erased, which in particular trivialises
the  cancellations for BPS type operators  for  $d\!=\!2$,  in contrast with the full consideration in \LMRS.
}

  \newsec{\bf 
   Alternative derivation of the  quasiclassical   OPE constant}
  
The derivation  follows and generalises  the   approach of \ZZ\ 
 in the Liouville theory,  so let us sketch the  main steps. 
 The  solutions of the classical  equations  of motion for the Liouville field  
 $$ \pd \bar{\pd} \varphi = \pi \mu b^2 e^{2\varphi}$$  
can be  recovered   from the solutions of the  second order chiral equation   (see, e.g. \GM\ and earlier references therein) 
\eqn\sv{
(\pd_z^2 +b^2 T_L(z))e^{-\varphi(z,\bz)}=0\,, \ \  \pd_{\bz} T_L=0
}
and its $\bar{T}_L(\bz)$ counterpart. 
The Liouville equation  of motion 
ensures  the conservation of the energy momentum tensor and vice versa. In the presence of (three) sources    the classical tensor is  determined
through the  limit $b^2\to 0$ of its normalised 4-point correlator with the three vertex operators
$e^{2\a_i \varphi\over b }$ of "heavy" charges  $\a_i = \eta_i/b$, or  
\eqn\tem{\eqalign{
&\hat{T}_L(z;z_a)=
\lim_{b\to 0} b^2 T(z;z_a)
= \sum_i({h_i\over (z-z_i)^2} +{\pd_i\over z-z_i}\log {1\over (z_{12})^{h_{12}^3}(z_{23})^{h_{23}^1}(z_{13})^{h_{13}^2}}) \cr
&= ({z_{12} z_{23}\over (z-z_2)^2 z_{13}})^{2}  \({h_1\over w^2 }+ {h_3\over (1-w)^2}+{h_{13}^2\over w(1-w)}\) = 
: (\pd_z w)^2\, \hat{T}_L(w)\,, \cr
}}
where $h_{ij}^k=h_i\!+\!h_j\!-\!h_k$ and 
\eqn\formdim{h_i=\eta_i(1-\eta_i)=\lim_{b\to 0} b^2 \alpha_i(Q-\a_i) \,, \ \    w= {(z-z_1)z_{23}\over (z-z_2)z_{13}}\,.
}
 Then the solution for $e^{-\varphi(z, \bz)}$ as a   function 
 of the coordinates $z_i, \bz_i$ of the sources is given, up to  a prefactor (determined by its  classical dimension $ -1/2=\lim_{b\to 0} \delta^L(-b/2) $),   by a 
 monodromy invariant diagonal combination of two solutions  of 
\eqn\blockL{
( \pd_w^2 + \hat{T}_L(w)) G^{\pm}(w) =0 \,.
 }
Equation  \sv\ 
 is  the classical version of the  BPZ equation,  
 resulting from the decoupling of a  level 2 singular vector.\foot{The   first derivative term in the quantum BPZ equation drops in this limit which justifies the definition  \tem.}
 The solution of \sv\  is  identified up to an overall constant with the classical limit of the  4-point function
of the    fundamental quantum vertex operator $e^{-b\phi}$ ($b\phi=\varphi$)
and the three arbitrary vertex operators,  
normalised by the 3-point function of these operators. 
Finally,  the solution for $\varphi(z;z_i)$   is   used to compute the saddle point action with sources which determines the semiclassical 3-point function \ZZ.
 
\medskip
\no 
\rb In the related to  a WZW model $AdS_3$ case, the BPZ equation  is replaced by the KZ equation  \KZ. The solution for the field  $V_{j=1/2}(z,\bz;x,\bar{ x})$ 
in \vert\  as  a function  of the coordinates  $\{z_i,\bz_i,x_i, \bx_i\}$ of the three sources 
is the classical limit of the corresponding 4-point function with one such vertex operator \ZF,\Te\ and three vertex operators \qvert\  of isospins  $j_i=-\eta_i/b^2$,  normalised by the corresponding 3-point function.  
For completeness let us  sketch  the  derivation of the KZ equation directly in the classical limit.  In the presence of three sources the chiral equation \chire\ 
for  \vert\  
becomes 
\eqn\kza{
(\pd_z - \hat{J}^a(z;z_i;x_i)t^a)V(z,x) =0\,,
}
with the current defined through the classical limit of its  4-point function normalised with the 3-point function
$<V_{j_1}V_{j_2}V_{j_3}>$
\eqn\clascur{\eqalign{
\hat{J}^a(z; z_i, x_i):& =
\lim_{b\to 0} {b^2\over <V_{j_1}V_{j_2}V_{j_3}>}\sum_i { t^a_i\over z-z_i}
<V_{j_1}V_{j_2}V_{j_3}>\cr
&=
x_{12}^{\eta_{12}^3}x_{23}^{\eta_{23}^1} x_{13}^{\eta_{13}^2}
\sum_i ({D^a_i(-\eta_i)\over z-z_i}-{D^a_i(-\eta_i)\over z-z_2}) \, x_{12}^{-\eta_{12}^3}x_{23}^{-\eta_{23}^1} x_{13}^{-\eta_{13}^2}\cr
&= : 
(\partial_z w)  \hat{J}^a(w;x_i) \,
}}
The $sl(2)$  generators are
 represented by the standard  differential operators $D_i^a(j_i)$ in $x_i$. In the last two lines we have used the Ward identities (projective invariance) with 
 $w$ defined in \formdim. 
The   current   $\hat{J}(w,x;x_i)= \hat{J}^a(w;x_i)t^a$ in the equation \kza\ 
becomes a differential operator with respect to the isospin projective invariant  $y$ when acting on the function
$\hat{V}(w,y)$ in 
\eqn\prefa{V(z,x;z_i,x_i)={|(x-x_2)^2 x_{13}|\over |x_{12} x_{23}|} \hat{V}(w,y)\,, \qquad y= {(x-x_1)x_{23}\over (x-x_2) x_{13}}\,.
}
Here  the dependence of $V$ and $\hat{V}$ on the  complex conjugated  variables  is suppressed . 
Since the Sugawara dimension  of $V=V_{j=1/2}$ vanishes in the limit $b\to 0$, there is no $z$-dependent prefactor in \prefa. 

One  obtains  the equation (written for the chiral constituents of $V$) 
\eqn\kz{\eqalign{
\pd_w G(w,y)&=
-\(\eta_{13}^2({(y-w)^2\over w(w-1)} -1) +2\eta_1
({y-w \over w} +1) +2\eta_3({y-w\over w-1} +1)\)\pd_y  G(w,y)\cr
&+\({\eta_{13}^2 (y-w)\over w(w-1)} + {\eta_1\over w} +{\eta_3 \over w-1}\)  G(w,y)\,.
}}

\no
\rb  The equation  \kz\  is equivalent  to a pair of differential  equations    for the components of 
\eqn\vr{
G(w,y)=G_0(w)+(y-w)G_2(w) \,. 
}
In  matrix form \kz\  reads  for the vector $G=(G_0,G_2)^t$
\eqn\maf{{d\over d w} \, G(w) - \pmatrix{&{\eta_1\over w}+{\eta_3\over w-1}& 1-\eta_{123}\cr
&{\eta_{13}^2\over w(w-1)} &-({\eta_1\over w}+{\eta_3\over w-1})}G(w)=0  \,. 
}
The component $G_0(w)=G(w,y=w)$ satisfies the second order equation  \blockL,  with the Liouville classical energy-momentum  tensor $\hat{T}_L(w)$.\foot{ I.e., with  the "classical"  scaling dimensions  given as in \formdim\ by $h_i=\eta_i(1-\eta_i)$ , while in the WZW classical stress tensor,  which we do not exploit here, they are given by $(-\eta^2_i)$.}  Thus the problem is reduced to that in the Liouville case.
This is an illustration of the  Drinfeld-Sokolov 
 reduction: 
gauge transformation 
by a  lower triangular  group element  which preserves $G_0=G_0^{\rm gau}$ and  brings the matrix equation  \maf\ to the form
\eqn\maf{{d\over d w}  G^{\rm gau}(w)- \pmatrix{&0&1-\eta_{123}\cr
&- {\hat{T}_L(w)\over 1-\eta_{123}}&0}G^{\rm gau}(w)=0  \,.
}

Combining the left and right  solutions   one obtains  for  $V$
\eqn\ffbb{\eqalign{
& 
V(z,x;z_i,x_i)  
={|x-x_2|^2|x_{13}|\over |x_{23}|| x_{12}
|N_{WZW}} (|G^{(+)}(w,y)|^2 +N_{WZW}^2 |G^{(-)}(w,y)|^2)\,, 
}}
 where 
 \eqn\fb{G^{(\pm)}(w,y)  = G_0^{(\pm)}(w) +(y-w) G_2^{(\pm)}(w)=G_1^{(\pm)}(w) + y\, G_2^{(\pm)}(w)
}
and the two solutions are given explicitly as

\eqn\solwz{\eqalign{
&G^{(+)}(w,y) =w^{\eta_1} (1-w)^{\eta_3} \times \cr
&\({}_2F_1(\eta_{13}^2, -1+\eta_{123}, 2\eta_1;w)-(y-w) {\eta_{13}^2\over 2\eta_1}{}_2F_1(1+\eta_{13}^2, \eta_{123}, 1+2\eta_1;w)\)
\cr
&= {w^{\eta_1} (1-w)^{\eta_3} \over  B(\eta_{12}^3, \eta_{13}^2)} 
 \int_1^{\infty} du\,  u^{\eta_{23}^1-1} (u-1)^{\eta_{12}^3-1}(u-w)^{1-\eta_{123}} {u-y\over u-w}\cr
&{}\cr
&
 G^{(-)}(w,y) 
 = w^{1-\eta_1} (1-w)^{\eta_3} \times
 \cr
  &\({1-\eta_{123}\over 1-2\eta_1}   \,  {}_2F_1(1-\eta_{12}^3, \eta_{23}^1, 2-2\eta_1;w) + {y-w\over w}\, 
 {}_2F_1(1-\eta_{12}^3, \eta_{23}^1, 1-2\eta_1;w)\)  \cr
 &= {w^{\eta_1} (1-w)^{\eta_3} \over B(1\!-\!\eta_{123}, \eta_{23}^1)} 
 \int_0^w du\,
 u^{\eta_{23}^1-1} (1-u)^{\eta_{12}^3-1}(w-u)^{1-\eta_{123}} {y-u\over w-u}
}}

\no
with  constants  expressed by beta functions $B(a,b)=\Gamma(a)\Gamma(b)/\Gamma(a+b)$. The last lines in \solwz\
correspond to the expansions in \fb\  in powers of $y$. 

The relative constant in \ffbb\  
\eqn\Nwzw{
{N^2_{WZW}}=  { \g( \eta_{23}^1) \g(2\eta_1)^2 \over   
 \g(\eta_{123}) \g( \eta_{12}^3 ) \g( \eta_{13}^2 )} =-{(1-2\eta_1)^2\over (\eta_{123}-1)^2}\, N^2_L 
}
is determined from the requirement of permutation invariance of the solution (crossing symmetry, or locality of the corresponding 
quantum 4-point correlator, see  \ZF\  for  the compact  WZW model, and   \Te\ for the non-compact  analog); 
$N_{WZW}$  is expressed by ratio of  products of fusing  matrices and $N_L$ in the r.h.s. of \Nwzw\ is the corresponding Liouville constant.
Note that  
 the basis  of contour integrals in \solwz\ transforms  (moving simultaneously  the pairs of space-time and isospin coordinates $(z_a,x_a)$)
  with  the   same braiding (fusing) matrices as in the Liouville  case: the shift from 
  $N_L$ to $N_{WZW}$ in \Nwzw\ is due to  different coefficient in front of the second contour integral in  \solwz\  when compared with the corresponding Liouville combination.\foot{The contour integrals have simple transformations described by linear combination of phases, which organise in $\sin$ - functions \DF. On the other hand these bases are  not normalised to $1$ when approaching one of the  three sources. When accounting for the additional constants the
fusing matrix elements 
are expressed by   $\Gamma$ - functions, leading to the $\gamma$ - functions in \Nwzw. 
The related 
 consideration  of  \JW\ 
seems to us not sufficiently clear at this point.  Recall that  
the gauge freedom in the braiding matrices  is  correlated  with  the normalisation of the chiral vertex operators $^{j_3}V_{j_2}^{j_1}$.} 
 The  Liouville solution itself  is reproduced identifying in \ffbb\  the  isospin variables with space time coordinates $x\to z\,, x_i\to z_i$.
 The overall constant in \ffbb\   is fixed by the equations  of motion, see below.

\medskip

\no
\rb Given the solution for $V$ we can  extract the expressions for the analogs of the matrix elements in the classical formulae \param, \vert,  i.e, $\phi$ and $\g,\bg$ as functions of $\{z_i,\bar{z}_i, x_i, \bar{x}_i\,, i=1,2,3\}$ and check the equations  of motion.
There is a certain arbitrariness in it since only $V$, not its ingredients, are monodromy invariant. We shall
 expand \ffbb\ in powers of $(x-x_1)$ in the vicinity of $x\sim x_1$.  More precisely,  introduce normalised chiral $u_i^{(\pm)}(w, x;x_i)$ as
\eqn\xexp{\eqalign{
&V =   |\psi_1^{(+)}+\psi_2^{(+)}|^2 + |\psi_1^{(-)}+\psi_2^{(-)}|^2
}}
\eqn\newdef{\eqalign{
\psi_1^{(\pm)}&= u_1^{(\pm)}(w,x;x_i)= {x-x_2\over x_{12}} ({x_{13}x_{12}\over x_{23}})^{1/2}{1\over (N_{WZW})^{\pm \ho}}(G_0^{(\pm)}(w)-wG_2^{(\pm)}(w))
\cr
\psi_2^{(\pm)}&=  (x-x_1) {x-x_2\over x_{12}} u_2^{(\pm)}(w, x;x_i)\cr 
& \Rightarrow u_2^{(\pm)}(w,x;x_i)=  {x_{12}\over x-x_2 } 
({x_{13}x_{12}\over x_{23} })^{-1/2}{1\over (N_{WZW})^{\pm \ho}}G_2^{(\pm)}(w)
}}
where  $G_0^{(\pm)}(w)\,, G_2^{(\pm)}(w)$ can be read from  \solwz, \vr. 
We then identify,   taking $u_i^{(\pm)}:=u_i^{(\pm)}(w, x=x_1; x_i)$ 
\eqn\sols{\eqalign{
&X_{-1}+X_3=e^\phi= \sum_{\pm} |u_2^{(\pm)} |^2 \,, \ \cr
&X=-X_1+i  X_2 = \g e^\phi=-\sum_{\pm}u_1^{(\pm)} \bar{u}_2^{(\pm)}\,, \ \bar{X}=\bg e^{\phi}= -\sum_{\pm}\bu_1^{(\pm)} {u}_2^{(\pm)} \cr
&X_{-1}-X_3 =e^{-\phi} +|\g|^2 e^{\phi}= \sum_{\pm} |u_1^{(\pm)} |^2
}}
where $\bar{u}_i$ is the complex  conjugate of $u_i$.  Or, in  a matrix form  for the  group element \param\
we have  (chiral factorisation)
 \eqn\solsa{
g(X)= h h^+\,, h=\pmatrix{u_1^{(+)} &u_1^{(-)} \cr u_2^{(+)} &u_2^{(-)} }\,.
}
The last equality in \sols\ (or, equivalently, the validity  of \hth)   requires that
\eqn\topr{
|u_1^{(+)}  u_2^{(-)}- u_1^{(-)} u_2^{(+)} |^2=|{\rm det}\, h|^2=1\,.
}
This is checked to hold true and more precisely 
 \eqn\solsb{
 u_1^{(+)}  u_2^{(-)} -u_1^{(-)}  u_2^{(+)} 
=1\,.
}
To prove \solsb\  one has to use the KZ equation   to express $G_2^{(\pm)}$ in terms of  $G_0^{(\pm)}$ and their  derivatives. The evaluation of the above  difference is then reduced to 
  the computation of the  Wronskian  
  of the two normalised  to $1$ solutions
in the Liouville case, which is a constant due to \blockL. 

Using once again \topr\ one checks  that the classical equations  of motion \eqm\ are indeed satisfied by \sols; this in particular fixes the overall constant in \ffbb. 

\medskip
\no
\rb The check of the  equations  of motion is done for $z,\bz$ far from the locations of the sources.
Let us  now look at the behaviour  of the solutions near one  of the sources when $z\to z_1\,, x \to x_1$. 
For $b\to 0$
one has 
$\d^{Su}(j)+\d^{Su}(1/2) -\d^{Su}(j\pm 1/2)\to 
\mp \eta\,.$ 

The leading contribution in  the fusion $j \to j+1/2$ is given by the first vector component of the first solution, namely  (taking here only the chiral factors)
\eqn\ftpl{\psi_1^{(+)} \sim  u_1^{(+)}
\sim{1\over \sqrt{N_{WZW}}}{
1\over (z-z_1)^{-\eta_1}} ({ z_{23} \over z_{13}  z_{12} })^{\eta_1} ({x_{13} x_{12}\over x_{23}})^{{1\over 2}}\,,
}
 while the 
second vector component  describes a descendant with respect of the finite subalgebra
\eqn\stpl{
\psi_2^{(+)} \sim 
(x-x_1)u_2^{(+)}
\sim  -{\eta_{13}^2 \over 2\eta_1}  {1\over \sqrt{N_{WZW}}}{
x-x_1\over (z-z_1)^{-\eta_1}} ({ z_{23} \over z_{13}  z_{12} })^{\eta_1} ({x_{13} x_{12}\over x_{23}})^{-{1\over 2}}\,.
}
For the contribution of $j \to j-1/2$ the leading term is given by the second vector component of the second solution
\eqn\lead{
\psi_2^{(-)}\sim (x-x_1)u_2^{(-)}
\sim \sqrt{N_{WZW}}{(x-x_1) \over (z-z_1)^{\eta_1}} ({ z_{23} \over z_{13}  z_{12} })^{-\eta_1} ({x_{13} x_{12}\over x_{23}})^{-{1\over 2}}\,,
}
while the first vector component  reads
\eqn\ftmn{
\psi_1^{(-)} \sim   u_1^{(-)}
 \sim 
-  {\eta_{23}^1\over 1-2\eta_1}   \sqrt{N_{WZW}}{1 
\over (z-z_1)^{\eta_1-1}} ({ z_{23} \over z_{13}  z_{12} })^{1-\eta_1} ({x_{13} x_{12}\over x_{23}})^{1\over 2}\,.
}
The behaviour  of the solutions for $z \sim z_2, $ or $z\sim z_3$ is described  permuting
$(1,2,3) \to (2,3,1) $, or $(1,2,3) \to (3,1,2)$.

From \ftpl\ and \lead\  one has that near $z\sim z_1\,, x\sim x_1$ 
\eqn\leadrep{
 \log |\psi_1^{(+)}|^2 \sim  - (\log |\psi_2^{(-)}|^2 - \log|x-x_1|^2)\sim -\log |u_2^{(-)}|^2\,.
}
The solution in the r.h.s of \leadrep\ gives  for $\eta_1>0$ the leading contribution 
to $\phi$,  defined  by the first equality in \sols, i.e.,  
 near the  source  
\eqn\adda{\eqalign{
&2\phi\sim 2\phi^{(-)}= 2\log |\psi_2^{(-)}|^2 - 2\log|x-x_1|^2  
 \sim  - 2\eta_1\log|z-z_1|^2 
+ X_1\,, 
\cr
}}
where
\eqn\Xone{ 
 X_1=  2 \eta_1 \log | {z_{13}  z_{12} \over  z_{23} }|^2 -  \log |{x_{13} x_{12}\over x_{23}}|^2 -
 \log {\g(\eta_{123}) \g( \eta_{12}^3 ) \g( \eta_{13}^2 )\over   \g( \eta_{23}^1) \g(2\eta_1)^2} \,.
}
Due to the symmetry of the classical solution analogous formulae hold in the vicinity of all three  singular points. 
We now add to the classical action  $S^{(cl)}=b^2 S_{AdS}^{'}$    terms which  account for the  three vertex insertions 
\eqn\shift{\eqalign{
\hat{S}^{(cl)}&= S^{(cl)}+ S^{(src)}, \cr
S^{(src)}&= - \int d^2 z \sum_i \delta^2(z-z_i, \bz-\bz_i) (2\eta_i\phi^{(-)}(z,\bz)
+  2\eta_i^2 \log|z-z_i|^2)\cr
&=-\sum_i \eta_i X_i \,.   
}}
Here it is  assumed that the integration in the first term  $S^{(cl)}$ is on  the  Riemann sphere with the points of insertion
of the 
three sources excluded.  The action \shift\ is regularised, the logarithmic  singularities  
 compensate those in  the
classical solution.  

At the saddle point of $S^{(cl)}=S^{(cl)}[\phi, \g,\bg] $, i.e., on a  solution of the classical equations,  
only the second term in  \shift\   contributes to the derivative of the full action with respect to any of 
the charges $\eta_i$. Thus  one obtains 
\eqn\syst{
{\pd\over \pd \eta_i} \hat{S}^{(cl)}=-X_i\,, i=1,2,3.
}  
This set of equations   integrates 
to 
\eqn\systs{\eqalign{
{ \hat{S}^{(cl)}
 \over b^2} 
  &=
   \sum_{i\ne j\ne k\ne i}( \d_{ij}^k \log |z_{ij}|^2- j_{ij}^k\log |x_{ij}|^2)\cr & +
{1\over b^2}(F(\eta_{123})+\sum_i F(\eta_{123}-2\eta_i)- \sum_i  F(2\eta_i) -F(0))\,, 
}}
\eqn\defdi{
{\rm where} \ \ \d_{ik}^l=\d_i\!+\!\d_k\!-\!\d_l\,, \ \ \delta_i  
=-{\eta_i^2\over b^2}\,, \ 
j_{ik}^l =  j_i\!+\!j_k\!-\!j_l\,, \ \ j_i=-{\eta_i\over b^2}\,. 
}
From the r.h.s. of \systs\ one reproduces    the 3-point function with coefficient  as in \adst\   up to the dependence on the normalisation factor $\nu(b)$
\eqn\fin{
e^{-{\hat{S}^{(cl)}\over b^2}}=C^{(cl)}(\eta_1,\eta_2,\eta_3)
{ |x_{13}|^{2j_{13}^2 }|x_{12}|^{2j_{12}^3} |x_{23}|^{2j_{23}^1}\over |z_{13}|^{2\d_{13}^2}  |z_{12}|^{2\d_{12}^3}|z_{23}|^{2\d_{23}^1}}\,.
}
Since $F(x)=F(1- x)$  the choice $F(0)=F(1)$  for the arbitrary  integration constant ensures that  $C^{(cl)}=1$ for $\sum_i \eta_i=1$,  a charge conservation condition, corresponding to absence of screening charges in the quantum case. 
\medskip

\no
\rb The semiclassical 3-point  contribution for $S^3$ is computed in precisely the same way, exploiting the solutions of  the classical 
KZ equation  in the compact WZW model ($b^2\!\to\!-b^2$); see, e.g.,  \HMW\ for the analogous consideration for the related 
$c<1$ Virasoro theory.  In the full 
  superstring theory 
    the  interrelation of the $AdS_3$ and $S^3$ contributions to the quantum 3-point correlator  of BPS  type  fields 
  has been described  in \GabKir, \DalPak.

\newsec{
Discussion  }

The   difference  with the  Ad$S_2$ result  is reduced  essentially to the function $\a \tilde{h}(\a)$  in (9.4) of \JW\  vs $F(\a)$ in \Fint.  
Though formally $AdS_2$ is a special case of $AdS_3$ 
it does not possess the affine symmetry of the WZW model.
If  at all legitimate to compare the semiclassics of 
these two different models, 
  the disagreement  might be also related to a  different normalisation  of the 3-point vertex itself,  see the comment in  footnote 5 above. 
 In both cases one analyses the solutions of a $2\times 2$  matrix differential equation with the same qualitative asymptotics around the singular points. Yet   
  we work here with explicit  full solutions of explicitly given 
 equation, so that there is a full   control on  the  coefficients in the asymptotics of the solutions in the vicinity of the sources. 

\medskip

 The 
 approach   followed here
can be extended in principle to the  computation of the WZW $AdS_5$, respectively $S^5$, contributions to the 3-point OPE constant of three scalar operators.  
Note that extrapolation   from the 
formulae computed   in the supergravity approximation  \FMMR, which can   be  interpreted as   "light charge"  limits of  the
(unknown) quantum expressions (cf.  \light),  suggests  that the  quantum formula for the   AdS$_{2n+1}$ 3-point scalar constant  for $n=2$ might be  
very close to the $n=1$
 expression in \adst\  -  with the shift  $-b$  in the first factor in the denominator replaced by $ - n  b=-2 b$.  If so, it  would  not  affect the  heavy charge limit of the constant, up to field renormalisations. 
 \medskip
 The   WZW $AdS_3 $  semiclassical OPE constants 
    could hardly be expected  to provide  an   approximation  of the corresponding scalar semiclassical OPEs 
 in the standard $AdS_5$ sigma model; this may  explain the difference with  the $AdS_2$ result in \JW. 
 On the other hand 
the  supersymmetric $AdS_5 \times S^5$ model in  RR background is conformally  invariant  \MTs,  at least in perturbation theory (see also the discussion in \KalTs\ and \Wiegmann).
This in our view  justifies the  elaboration of  CFT techniques in  semiclassical considerations.
 
\bigskip

\no
{\bf Acknowledgements}
\medskip

\noindent
We thank  Romuald Janik,  Ivan Kostov and Tristan McLoughlin for useful comments. 
VBP acknowledges the    hospitality of  the Istituto Nazionale di Fisica Nucleare (INFN),
Sezione di Trieste, Italy, where this work  was started. This research is  
partially supported   by the  Bulgarian NSF grant {\it DO 02-257}.

\listrefs
\bye